\documentclass[
 aip, amsmath,amssymb,
 reprint,%
]{revtex4-1}

\usepackage{graphicx}
\usepackage{dcolumn}
\usepackage{bm}

\usepackage[utf8]{inputenc}
\usepackage[T1]{fontenc}
\usepackage{mathptmx}
\usepackage{etoolbox}

\usepackage{hyperref}
\usepackage{xcolor}

\newcommand{\nn}{\nonumber}
\newcommand{\veps}{\varepsilon}

\makeatletter
\def\@email#1#2{%
 \endgroup
 \patchcmd{\titleblock@produce}
  {\frontmatter@RRAPformat}
  
{\frontmatter@RRAPformat{\produce@RRAP{*#1\href{mailto:#2}{#2}}}\
frontmatter@RRAPformat}
  {}{}
}%
\makeatother
\begin{document}

\preprint{AIP/123-QED}

\title[Liquid nucleation around charged particles in the vapor phase]{Liquid 
nucleation around charged particles in the vapor phase}
\author{Roni Kroll}
\author{Yoav Tsori}
 \email{tsori@bgu.ac.il}
\affiliation{ 
Department of Chemical Engineering, Ben-Gurion University of the 
Negev, Beersheba, Israel.
}%

\date{\today}

\begin{abstract}
We theoretically investigate the nucleation of liquid droplets from vapor in the 
presence of a charged spherical particle. Due to field gradients, sufficiently 
close to the critical point of the vapor--gas system, the charge destabilizes the 
vapor phase and initiates a phase transition. The fluid's free energy is 
described by the van der Waals expression augmented by electrostatic energy and 
a square-gradient term. We calculate the equilibrium density profile at 
arbitrary temperatures, particle charges, and vapor densities. In contrast 
to classical nucleation theory, here, both liquid and vapor phases are different 
from the bulk phases because they are spatially nonuniform. In addition, the 
theory applies to both sharp and diffuse interfaces and calculates the surface 
tension self-consistently. We find the composition profiles and integrate 
them to get the adsorption near the particle. We find that the adsorption changes 
discontinuously at a first-order phase transition line. This line becomes a 
second-order phase transition at high enough temperatures. We describe the 
transition point numerically and provide approximate analytical expressions for 
it. Similarly to prewetting, the adsorption diverges at the binodal phase 
boundary. We construct a phase diagram indicating changes in the binodal, 
spinodal, and critical temperature. It is shown that the field gradient enlarges the 
range of temperature and vapor density where liquid can nucleate.
\end{abstract}

\maketitle 

\section{INTRODUCTION}

Nucleation is a thermodynamic process that constitutes the initial step for many 
physical processes, such as vapor condensation, melting, and boiling. 
\cite{Coquerel_2014,Jones1999,Pruppacher1998,Martin2000} It is a localized 
phase transition process that occurs at metastable conditions, where the system 
has an energy barrier to overcome to reach its equilibrium state. Heterogeneous 
nucleation occurs in the presence of foreign objects, such as the walls of a 
vessel, dust particles, or other impurities. The presence of foreign particles 
reduces the nucleation energy barrier locally and, consequently, the 
supersaturation required for the nucleation. The rate of the critical nuclei 
creation is proportional to the exponential  of the ratio between the barrier 
and thermal energies. Heterogeneous nucleation can be relatively fast, and in 
many cases, it is the dominant mechanism. \cite{Turnbull_1950,Fletcher_1958}

A commonly used model for nucleation is the classical nucleation theory (CNT) 
based on the work of Volmer and Weber, Becker and D\"{o}ring and Frenkel. 
\cite{Volmer_1926,Becker_1935,Frenkel_1939} The CNT model describes the 
condensation of vapor to liquid nucleus. The nucleus, assumed to be spherical, 
is described by the macroscopic properties of the stable phase. The interface 
between the liquid and vapor phases is considered to be spherical with zero 
thickness, independent of the nucleus size. 

The Thomson model is an extension of CNT theory that includes the charge of the 
nucleating particle. In the Thomson model, the Gibbs transfer energy 
for the transfer of molecules from vapor to liquid around the charged particle 
is given by \cite{Laaksonen_1995}
\begin{align}
\Delta G=&-\frac{4\pi R_{i}^{3}}{3v}k_{B}T\ln(P/P^{\ast})+4\pi 
R_{i}^{2}\gamma\nonumber\\
&+\frac{q^{2}}{8\pi\varepsilon_{0}}\left(\frac{1}{\varepsilon_{1}}-\frac{1}{\varepsilon_{2}
}\right)\left(\frac{1}{R_{i}}-\frac{1}{R}\right).
\label{Gibbs_Transfer}
\end{align}
The first term represents a bulk free energy: $4\pi R_{i}^{3}/3v$ is the number of 
molecules in a liquid nucleus with radius $R_i$, where $v$ is the volume of a 
single molecule, and $P/P^*$ is the vapor supersaturation ratio. The second term 
is a surface energy, where $\gamma$ is the surface tension between liquid 
and vapor. The third term is the electrostatic energy of a spherical particle of 
radius $R$ and charge $q$ purely embedded in a dielectric fluid. $\veps_1$ and 
$\veps_2$ are the relative dielectric constants of pure gas and liquid, 
respectively, and $\veps_0$ is the permittivity of the vacuum.

\begin{figure}[h!]
\includegraphics[width=0.48\textwidth,bb=110 304 470 483,clip]{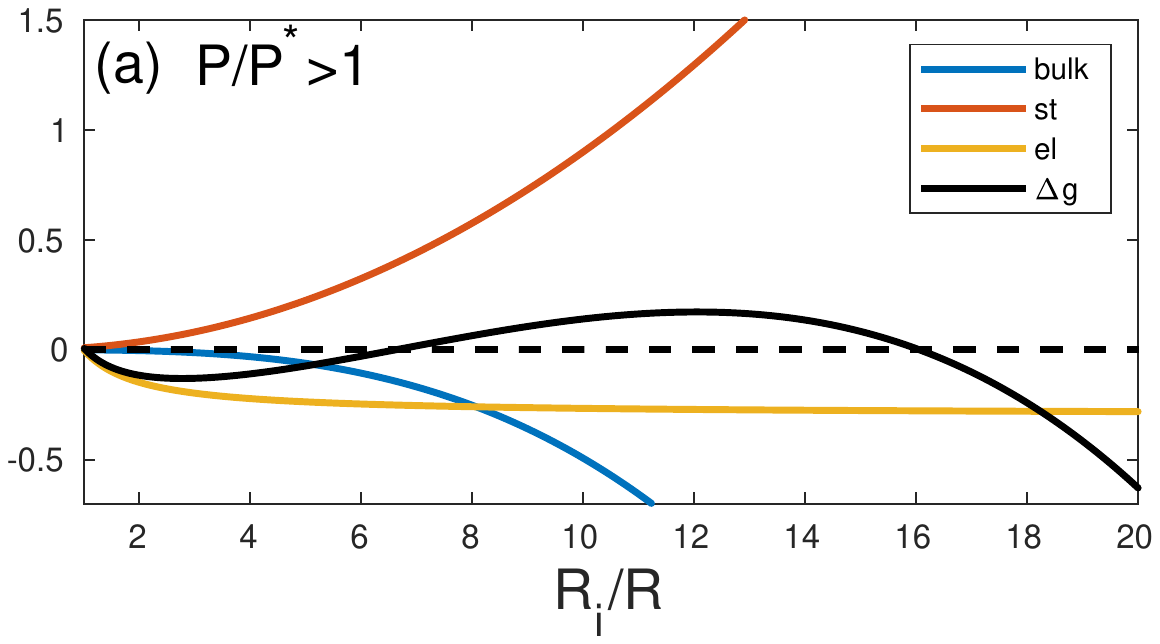}
\includegraphics[width=0.48\textwidth,bb=110 304 470 483,clip]{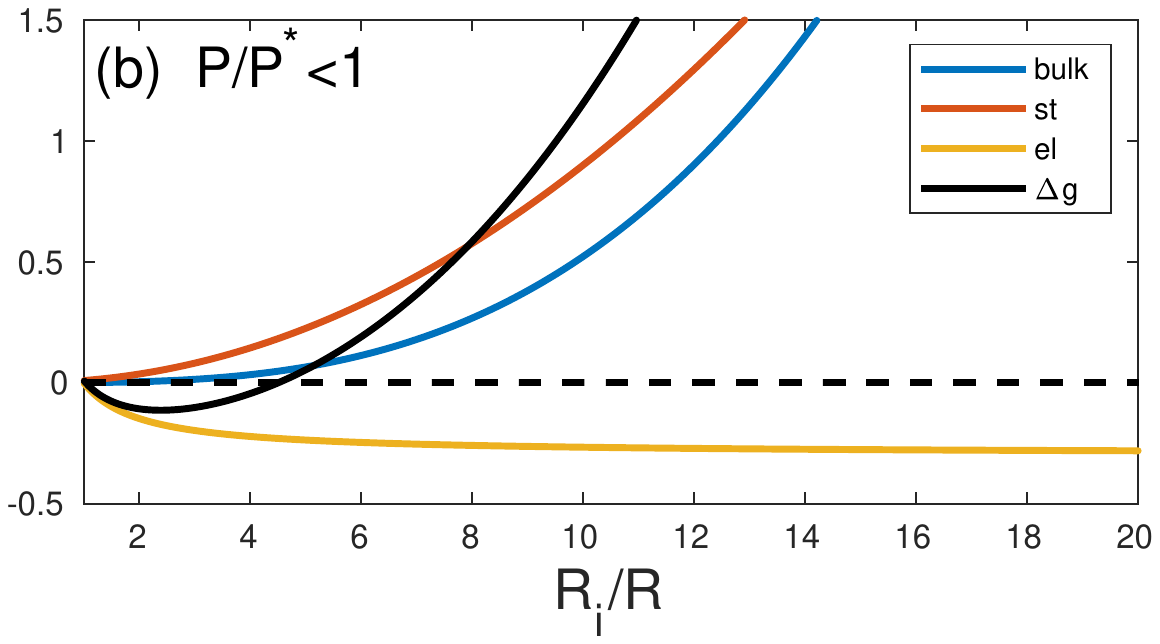}
\label{Thomson_b}\\
\includegraphics[width=0.48\textwidth,bb=110 304 470 475,clip]{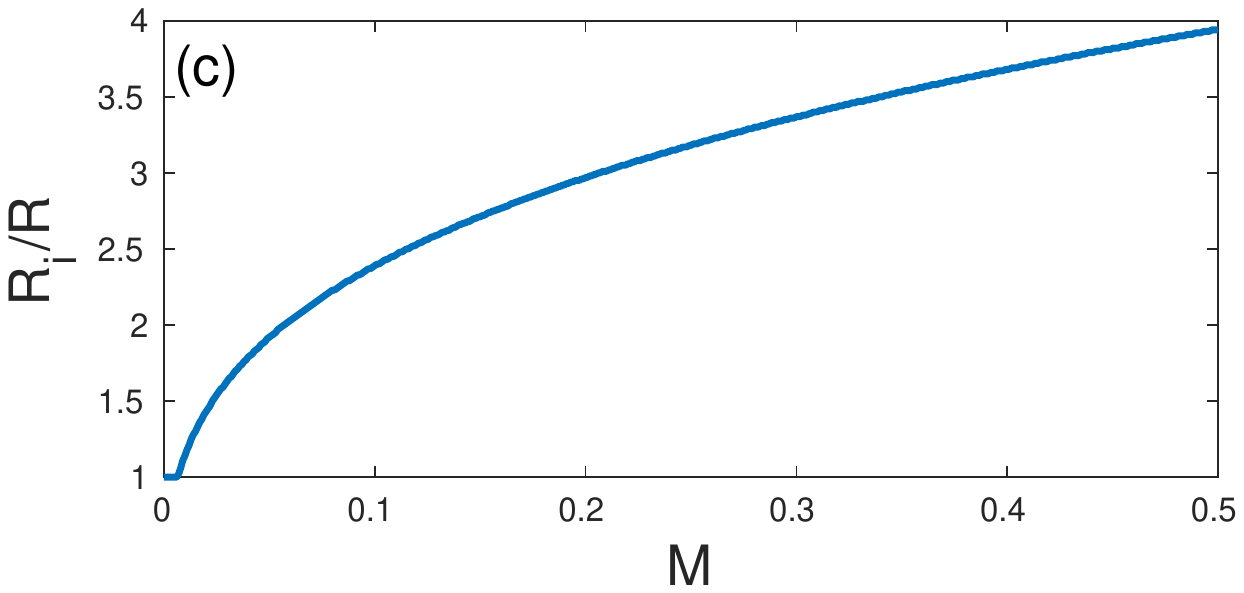}
\label{Thomson_c}
\caption{Different contributions to the energy in the Thomson model Eq. 
(\ref{Gibbs_Transfer}). (a) For a supersaturated system, 
$\phi_0=0.856$ and $M=0.1$. 
Legend labels ``bulk'', ``st'', ``el'', and ``$\Delta g$'' represent the mixing, surface 
tension, electrostatic, and total Gibbs energies, respectively.
(b) The same as in (a) but for 
an unsaturated system $\phi_0=0.850$. (c) The nucleus radius 
as a function of $M\equiv \sigma^2/2P_c\veps_0$, where $\sigma$ is the surface 
charge density of the charged particle. We used $\phi^*=0.853$, $T/T_c=0.995$, and 
$\gamma/(RP_c)=3\times 10^{-3}$.}
\label{Thomson}
\end{figure}
The nucleation process, in the absence of external forces, happens in 
supersaturated systems $(P/P^*>1)$. Once the conditions for nucleation exist, 
the Thomson model predicts the infinite growth of the radius as the Gibbs transfer 
energy is decreasing as a function of $R_i$. Figure \ref{Thomson}(a) shows the 
three energy contributions of the Thomson model in the metastable region of the 
phase diagram and their sum including the electrostatic term (black curve). A 
local minimum occurs at a finite radius $R_i$, but the most stable nucleus size 
is infinite. Figure \ref{Thomson}(b) shows the same energy contributions when the 
bulk conditions are not metastable, that is, in the stable vapor phase. In this 
case, the bulk term, $\sim R_{i}^{3}$, is positive, and due to the electrostatic term, 
the global minimum occurs at a finite nucleus radius $R_i$. The size of this 
nucleus vs $\sigma^2/2P_c\veps_0$ is displayed in Fig. \ref{Thomson}(c), where 
$\sigma$ is the surface charge density of the charged particle.

Experimental work in different fields such as polymer crystallization and 
atmospheric science confirmed that electrostatic interactions promote nucleation. 
\cite{Tosi_2008,Gamero_2002} This article generalizes the Thomson nucleation 
model. Near charged particles, field gradients lead to coupling between the 
electric field and the density of the fluid. The dielectric fluid is attracted 
by a dielectrophoretic force to the charged surface, leading to an increase in 
the density. In return, the density affects the electric field through the 
change in the dielectric constant. \cite{Tsori_nature} Below, we demonstrate how 
this phenomenon provokes the phase transition and enlarges the range of 
temperatures and densities where vapor and liquid coexist. We use the van der 
Waals mean-field energy supplemented by square-gradient theory. We extremize the 
free energy and solve the Euler--Lagrange equations to obtain the 
thermodynamically stable solutions. 

\section {MODEL} 

We investigate the liquid--vapor pure component system by using the classical van der 
Waals mean-field model. The fluid is characterized by its temperature $T$ and 
its number density $\rho$. The fluid surrounds a perfect solid sphere with 
radius $R$ that is uniformly charged with surface charge density $\sigma$. 
The Helmholtz free energy of a one component van der Waals fluid around a 
charged particle is the integral over space of the sum energy densities. 
Smooth density profiles, not considered by the CNT model, are allowed here by 
inclusion of a square-gradient term, \cite{Rowlinson_Widom_1982} 
\begin{equation}
F=\int{\left[\frac{1}{2}c^{2}\vert \nabla\rho \vert^{2}+f_{\rm {vdw}}+f_{\rm{es}}\right]dr}.
\end{equation}
Here, $c$ is a constant and $f_{\rm vdw}$ is the van der 
Waals free energy density given by \cite{barrat_hansen_2003}
\begin{equation}
f_{\rm vdw}=k_{B}T\rho\left[\log(\rho\Lambda^{3})-1-\log(1-\rho b)\right]-a\rho^{2}.
\end{equation}
Here, $k_B$ is the Boltzmann constant, $T$ is the temperature, and $\Lambda$ is 
the thermal de Broglie wavelength. The parameters $a$ and $b$ are positive 
constants where $a$ accounts for the attractive forces between the molecules and 
$b$ represents the excluded volume of the molecules. $a$ and $b$ are 
related to the critical temperature $T_c$ and density $\rho_c$ by 
$(T_c,\rho_c)=(8a/(27k_Bb),(3b)^{-1})$.
From the van der Waals equation of state $(P+a\rho^2)(1-b\rho)=\rho k_BT$, the 
value of the critical pressure is found to be $P_c=a/(27b^2)$.

The electrostatic energy density of a perfect dielectric fluid is 
\begin{equation} 
f_{\rm es}=\frac{1}{2}\veps_{0}\veps {\bf E}^2, 
\end{equation} 
where ${\bf E}=-\nabla\psi$ is the electric field, $\psi$ is the electrostatic 
potential, and $\veps(\rho)$ is the relative dielectric constant. The system is 
open and has a vapor phase with constant density far from the particle surface. 
The thermodynamically stable solutions are minimizers of the grand canonical 
energy $\Omega=F-\mu N$, where $\mu$ is the chemical potential and $N$ is the 
number of fluid molecules. Extremization of $\Omega$ with respect to density and 
electrostatic potential yields two coupled Euler--Lagrange equations,
\begin{eqnarray}
\label{Euler-Lagrange1}
\frac{\delta \Omega}{\delta \rho}&=&-c^2\nabla^2\rho+\frac{\partial f_{\rm 
vdw}}{\partial\rho}-\frac{1}{2}\veps_0\frac{d\veps}{d\rho}|\nabla\psi|^2-\mu=0,
\\\label{poisson}
\frac{\delta \Omega}{\delta 
\psi}&=&\nabla\cdot\left(\veps_0\veps(\rho)\nabla\psi\right)=0.
\end{eqnarray}
Equation (\ref{poisson}) represents the Laplace equation. The spherical symmetry allows us 
to describe the system in one dimension, with variations only in the $r$ 
coordinate; hence, the solution for the electric field, given by  ${\bf 
E}=\sigma R^2/r^2 \veps_0\veps(\rho){\bf \hat{r}}$, is decaying as ${\sim} r^{-2}$. 
We use $\phi\equiv\rho/\rho_c$ as the reduced density and substitute the 
electric field into Eq. (\ref{Euler-Lagrange1}) to get
\begin{equation}
-\tilde{c}^2\tilde{\nabla}^2\phi+\frac{\partial \tilde{f}_{\rm 
vdw}}{\partial\phi}-M\frac{d\veps/d\phi}{\veps(\phi)^2}\tilde{r}^{-4}-\tilde{\mu
}=0,
\label{equilibrium}
\end{equation}
where the tilde sign indicates reduced quantities:  $\tilde{f}_{\rm vdw}=f_{\rm 
vdw}/P_c$, $\tilde{r}=r/R$, $\tilde{c}^2=c^2\rho_c^2/P_cR^2$ and 
$\tilde{\mu}=\mu\rho_c/P_c$. In addition, we define
\begin{equation}
M=\sigma^2/2P_c\veps_0
\end{equation}
as the dimensionless electrostatic energy.
The boundary conditions for the equation are as follows:
\begin{equation}
 \begin{aligned}
  \tilde{r}=1,&\qquad -\hat{n}\cdot(-\tilde{c}^{2}{\nabla}\phi)=0,\\
  \tilde{r}=\infty,&\qquad\phi=\phi_{0},
 \end{aligned}
\end{equation}
where $\hat{n}$ is a unit vector perpendicular to the surface of the particle.
The first boundary condition means zero flux on the surface of the particle, 
while the second boundary enforces the bulk reservoir density $\phi_0$ at 
infinity. 

The local dielectric constant $\veps(\phi)$ depends on the local value of the 
density and on the dielectric constants of the pure phases, $\veps_1$ (gas) and 
$\veps_2$ (liquid). We assume a linear relation between the local dielectric 
constant and the fluid density, $\veps=\veps_1+\Delta\veps\phi/3$, where 
$\Delta\veps=\veps_2-\veps_1$ is a constant (recall that $0\leq \phi\leq 3$). In 
thermodynamic equilibrium, the chemical potential is constant everywhere. In the 
grand canonical ensemble, it is found from $\tilde{\mu}=\partial \tilde{f}_{\rm vdw}(\phi_0)/\partial \phi$ with  $\tilde{r}\rightarrow\infty$ where the 
electric field tends to zero and the density tends to the vapor reservoir 
density $\phi_0$.

\section {RESULTS AND DISCUSSION}
\subsection {Linearization of $\phi$}
We focus on a charged particle in the presence of a stable vapor phase. 
For small enough particle potentials, the fluid's density will not change 
sharply in space, and in this case one can linearize the Euler--Lagrange 
equations around the bulk value of $\phi$.
Substitution of $\phi=\phi_0+\delta \phi$, with $\delta\phi<<\phi_0$, into Eq. 
(\ref{equilibrium}) leads to a linear differential equation for $\delta \phi$,
\begin{eqnarray}
\tilde{\nabla}^2\delta\phi-A\delta\phi+B\tilde{r}^{-4}=0,
\label{Linearized_equation}
\end{eqnarray}
where $A$ and $B$ are given by 
$A=\left(\frac{24T/T_c}{\phi_0(3-\phi_0)^2}-6\right)/\tilde{c}^2$ and 
$B=M \Delta\veps/(3\veps^2\tilde{c}^2)$.
The general solution for Eq. (\ref{Linearized_equation}) is given by
\begin{eqnarray}
\delta\phi&=&C 
\frac{e^{-\sqrt{A}\tilde{r}}}{\tilde{r}}\\
&+&\frac{\sqrt{A}B}{4\tilde{r}}\left[-e^{
\sqrt{A}\tilde{r}}Ei\left(-\sqrt{A}\tilde{r}\right)+e^{-\sqrt{A}\tilde{r}}Ei
\left (\sqrt{A}\tilde{r}\right)\right]-\frac{B}{2\tilde{r}^2}\nn,
 \label{Linearized_solution}
\end{eqnarray}
where the first term represents the homogeneous solutions and the rest represents the 
particular solution. $Ei$ is the exponential integral defined as 
$Ei(x)=\int_{-\infty}^x (e^{t}/t) dt$.
The constant $C$ is found from the boundary condition $\partial \delta 
\phi/\partial \tilde{r}|_{\tilde{r}=1}=0$,
\begin{eqnarray}
C&=&-\frac{1}{4}\sqrt{A}B\left[\frac{\sqrt{A}-1}{\sqrt{A}+1}e^{2\sqrt{A}}Ei\left
( -\sqrt{A}\right)+Ei\left(\sqrt{A}\right)\right]\nn\\
&+&B\frac{e^{\sqrt{A}}}{\sqrt{A
}+1}.
\end{eqnarray}

\subsection {Nonlinear profiles in the sharp interface limit}

The linear profiles in Eq. (\ref{Linearized_solution}) are valid when the 
density changes are small, but when the electro-prewetting transition occurs and 
liquid wets the particle, this assumption falls, and one needs to solve Eq. 
(\ref{equilibrium}) numerically. The density profiles are calculated by setting 
the values of $\phi_0$, $M$, and $T/T_c$ and, then, finding the solutions 
$\phi(\tilde{r})$  for all $\tilde{r}$. When $\tilde{c}=0$, for small or large 
values of $\tilde{r}$, there is only one solution. However, if $M$ is large enough 
for intermediate values of $\tilde{r}$, there is more than one solution to the 
equation. At these $\tilde{r}$'s, we calculated the energy for each of the 
solutions and selected the equilibrium solution as the one with minimal energy.

\begin{figure}[h!]
 \includegraphics[width=0.45\textwidth,bb=90 230 490 
550,clip]{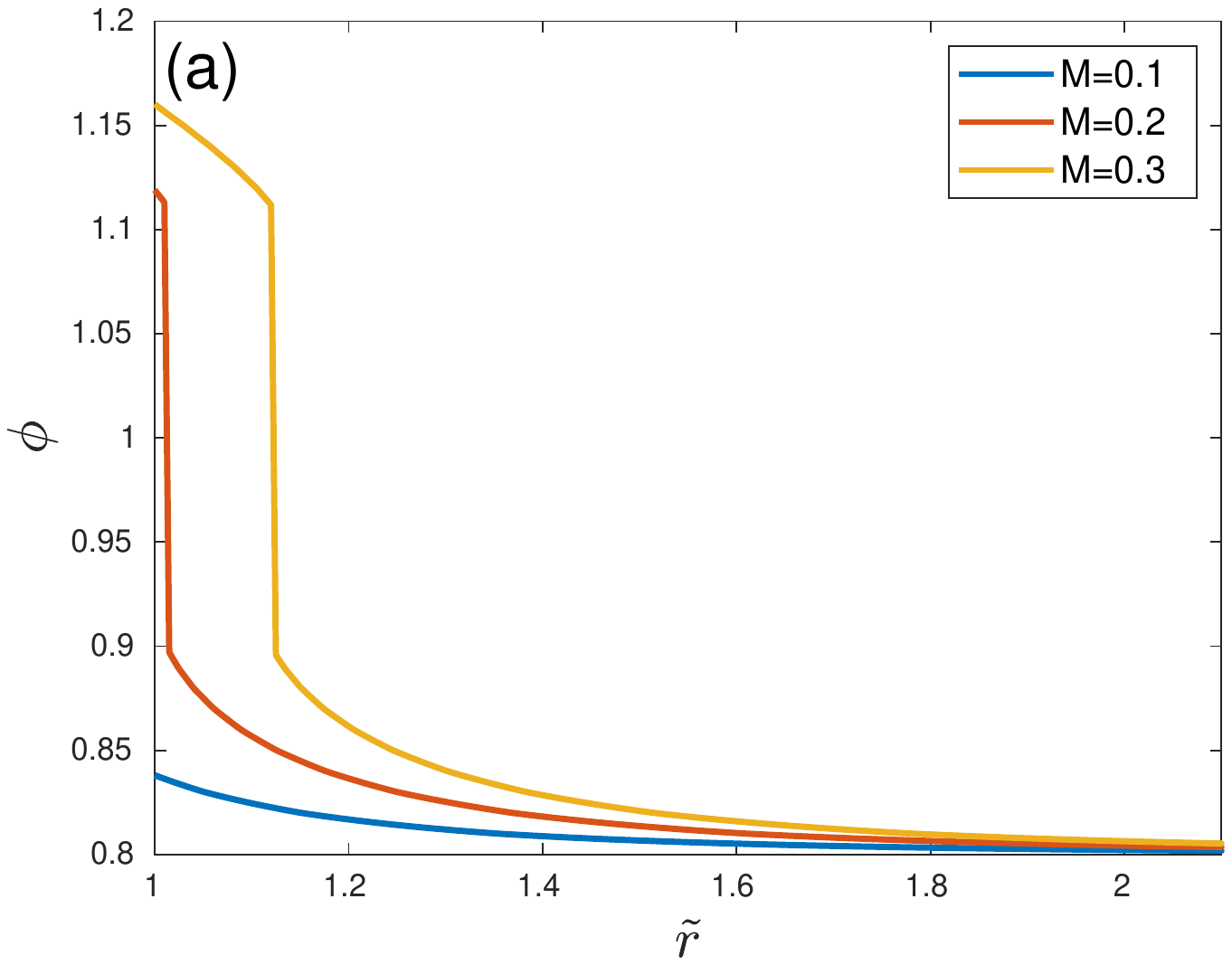}
 \includegraphics[width=0.45\textwidth,bb=90 230 490 550,clip]{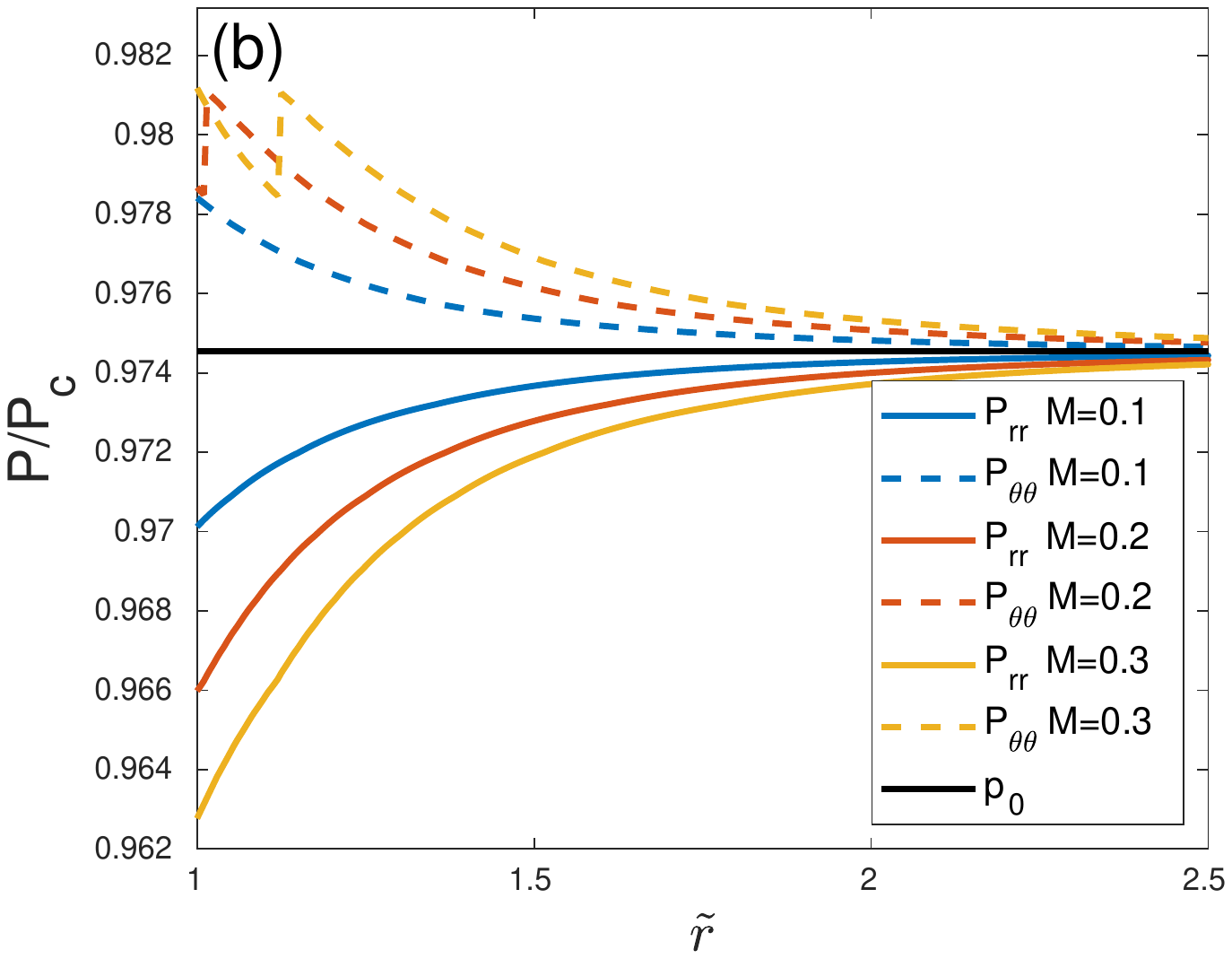}
\caption{(a) Equilibrium density profiles $\phi(\tilde{r})$ in 
the sharp interface limit, $\tilde{c}=0$, at a temperature of $T/T_c=0.995$ and 
bulk density of $\phi_0=0.8$. (b) Radial $p_{rr}$ and azimuthal 
$p_{\theta\theta}$ pressure profiles corresponding to the density profiles in 
(a). $\veps_1=1$ and $\veps_2=80$ in this and in all other figures.}
\label{density_profile}
\end{figure}

Fig. \ref{density_profile}(a) shows the numerically obtained density profiles in 
the sharp interface limit, $\tilde{c}=0$. At small values of $M$ (small particle 
charge), the density is smoothly varying at all values of $\tilde{r}$. At large 
enough values of $M$, a dense phase appears on the surface of the particle in 
coexistence with a vapor phase far from it. Note that both phases are spatially 
nonuniform and cannot be described by the bulk phases, liquid or vapor. The 
interface between the two phases exhibits a sharp jump in the density. We denote 
by $R_i$ the location of the vapor--liquid interface. Due to the 
dielectrophoretic force, an increase in $M$ ``draws'' more molecules to the 
region with strong electric field, resulting in a larger nucleus.

In Fig. \ref{density_profile}(b), we show the pressure profiles. 
When a fluid is under an electric field, stress develops. 
The stress tensor is given by \cite{Landau:712712}
\begin{eqnarray}
T_{ij}&=&-p_0(\phi,T)\delta_{ij}+\frac{1}{2}\veps 
E^2\left(-1+\frac{\phi}{\veps}\left(\frac{\partial\veps}{\partial\phi}\right)
\right)\delta_{ij}\nn\\
&+&\veps E_iE_j,
\end{eqnarray}
where $p_0$ is the bulk (zero charge) pressure given by 
$p_0=\phi_0\partial f_{\rm vdw}(\phi_0)/\partial \phi-f_{\rm 
vdw}(\phi_0)$. 
$\delta_{ij}$ is the Kronecker delta function. 
The diagonal pressure components in the $rr$ and $\theta\theta$ axis are
\begin{eqnarray}
p=p_0(\phi,T)-M\left[\pm\frac{1}{\veps}+\frac{\phi}{\veps^2}\left(\frac{
\partial\veps}
{\partial\phi}\right)\right]\tilde{r}^{-4},
\end{eqnarray}
where the (+) sign is for $p_{rr}$ and (--) sign for $p_{\theta\theta}$. The 
pressure profiles are discontinuous when a sharp interface is created. The 
discontinuity in $p_{rr}$ is  small.

\begin{figure}[h!]
\includegraphics[width=0.48\textwidth,bb=90 230 485 550,clip]{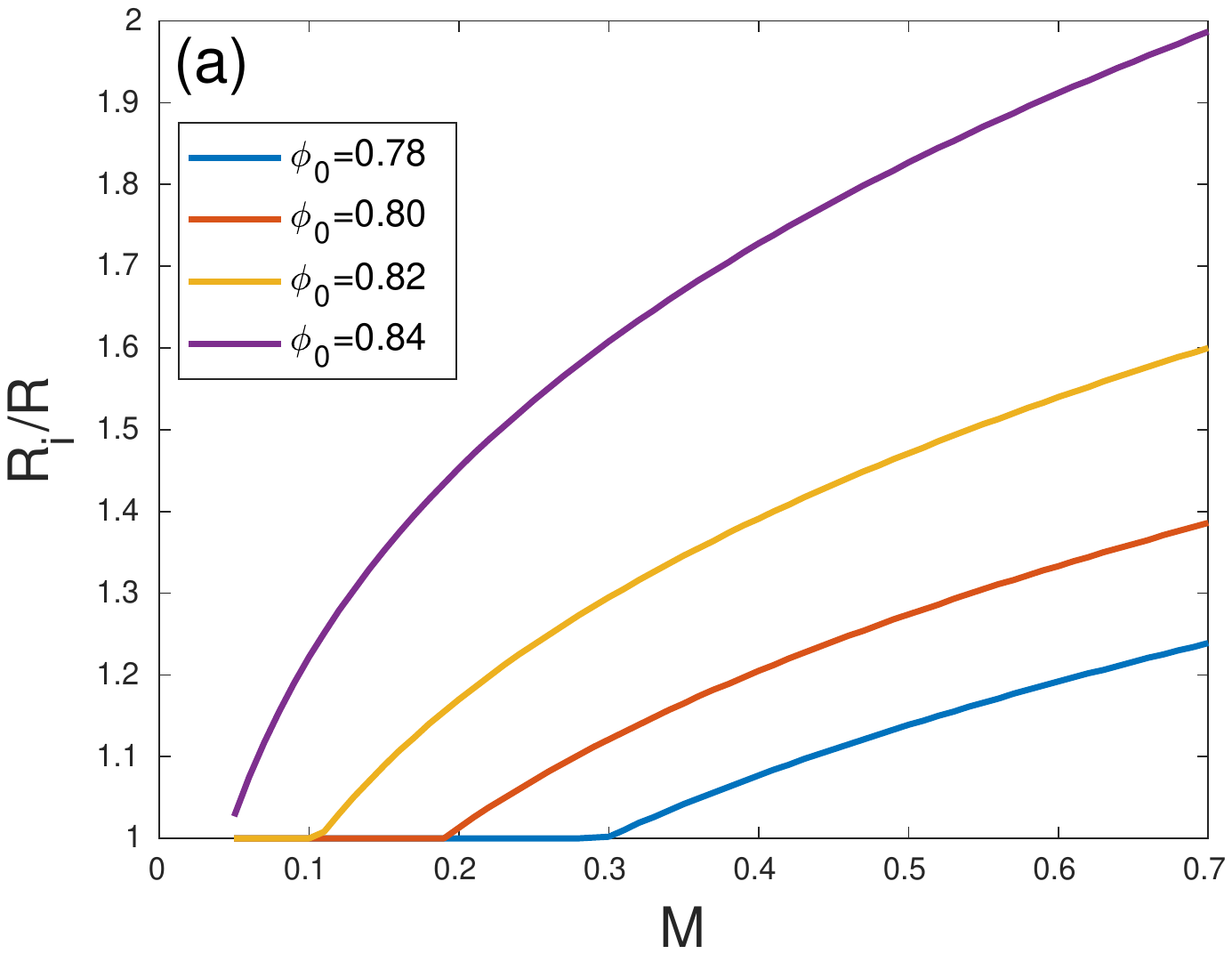}
\includegraphics[width=0.48\textwidth,bb=90 230 490  550,clip]{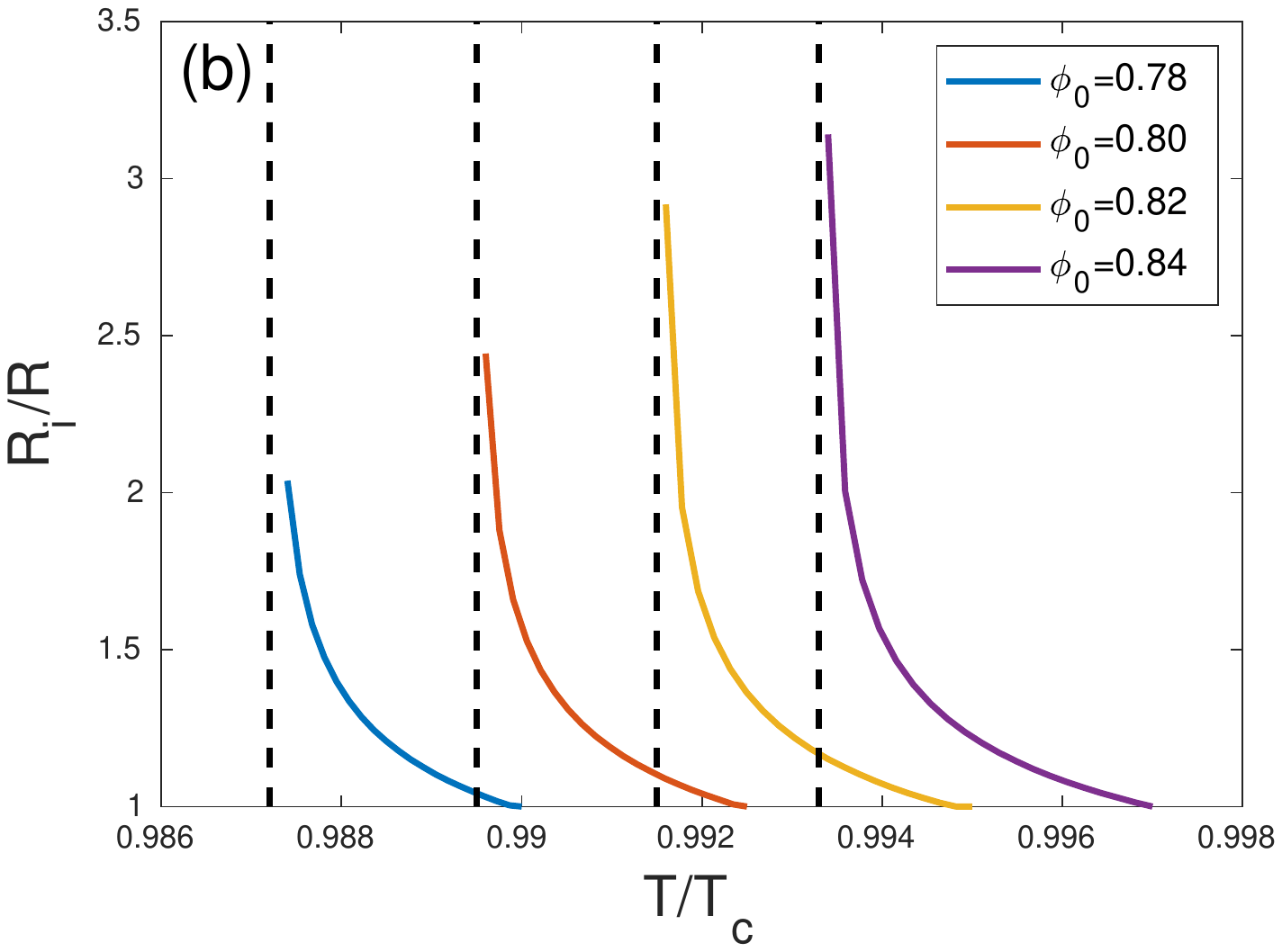}
\caption{(a) Location of the interface $R_i$ in the sharp interface limit, 
$\tilde{c}=0$, as a function of the scaled particle charge $M$ at fixed 
temperature given by $T/T_c=0.995$ for several values of the average 
composition $\phi_0$. (b) Similar to (a) but now $R_i$
as a function of $T/T_c$ at fixed $M=0.1$.}
\label{Radius_of_Drop_M}
\end{figure}

%
Figure \ref{Radius_of_Drop_M}(a) shows the finite 
radius $R_i$ of the nucleus as a function of $M$, obtained from the solutions 
of Eq. (\ref{equilibrium}) in the sharp interface limit. As the value of 
$\phi_0$ decreases, that is, $P/P^*$ decreases, a larger value of 
$M$ is needed for nucleation to occur. For the same $M$, a larger radius is 
obtained as the saturation ratio grows. 
In Fig. \ref{Radius_of_Drop_M}(b), we keep $M$ constant and vary the 
temperature. 
The radius of the drop $R_i$ increases when $T$ decreases. For each value of 
$\phi_0$, the lowest temperature shown is the binodal temperature, below 
which phase coexistence occurs even without the driving force of the electric 
charge. Below the binodal, the equilibrium nucleus size becomes infinite.

\subsection {Diffuse interface profiles}\label{Diffuse_interface}

Experimental data show that the nucleation rate in some cases deviate by several 
orders of magnitude from the CNT theory. \cite{Strey_1986,Laaksonen_2001} The 
source of the deviation is presumably the ``capillarity'' approximation, i.e., 
the assumption that the bulk surface tension of the nucleus equals the surface 
tension of a thin and flat interface. One popular approach to avoid these 
assumptions is to use the density functional theory (DFT), but it requires 
knowledge of intermolecular potentials. \cite{Oxtoby_1988,Alekseechkin_2018} 
Another approach, is the phenomenological diffuse interface theory, which has 
proven as reliable over many length-scales. \cite{Granasy_1993} We now release 
the assumption $\tilde{c}=0$ and solve the full equation Eq.~[(\ref{equilibrium})]
including the square-gradient term. This term is connected to changes in the 
density and adds the interfacial contribution to the free energy.   In spherical 
symmetry, the nonlinear static equation is solved in one dimension using finite 
elements and Newton--Raphson iterations. For numerical purposes, the maximum 
value of $\tilde{r}$ was taken as $\tilde{r}=10$, and indeed the gradient 
$|\phi'|$ was verified as negligibly small in these ``large'' distances.

\begin{figure}[h!]
\includegraphics[width=0.48\textwidth,bb=90 230 485 550,clip]{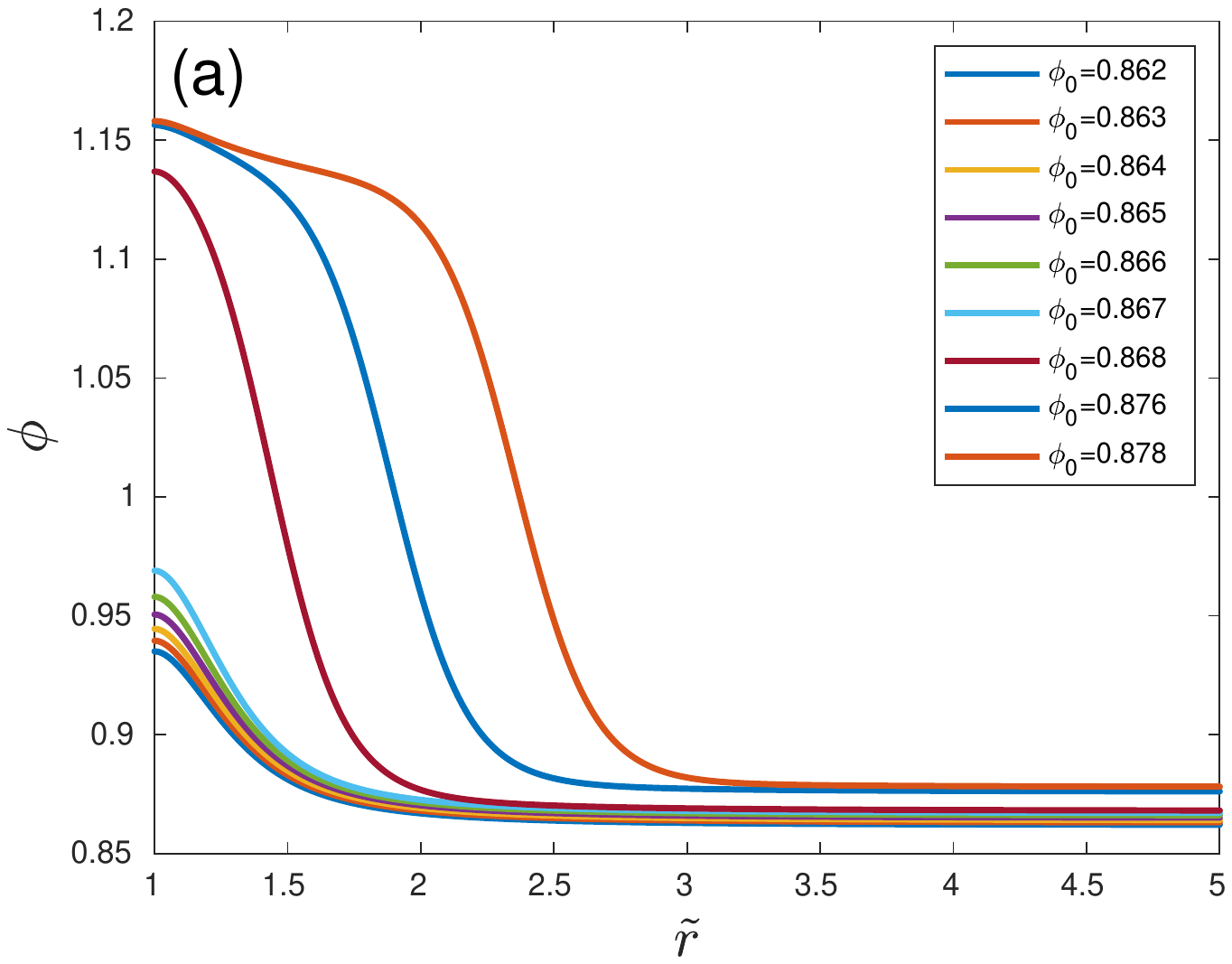}
\includegraphics[width=0.48\textwidth,bb=90 230 485 550,clip]{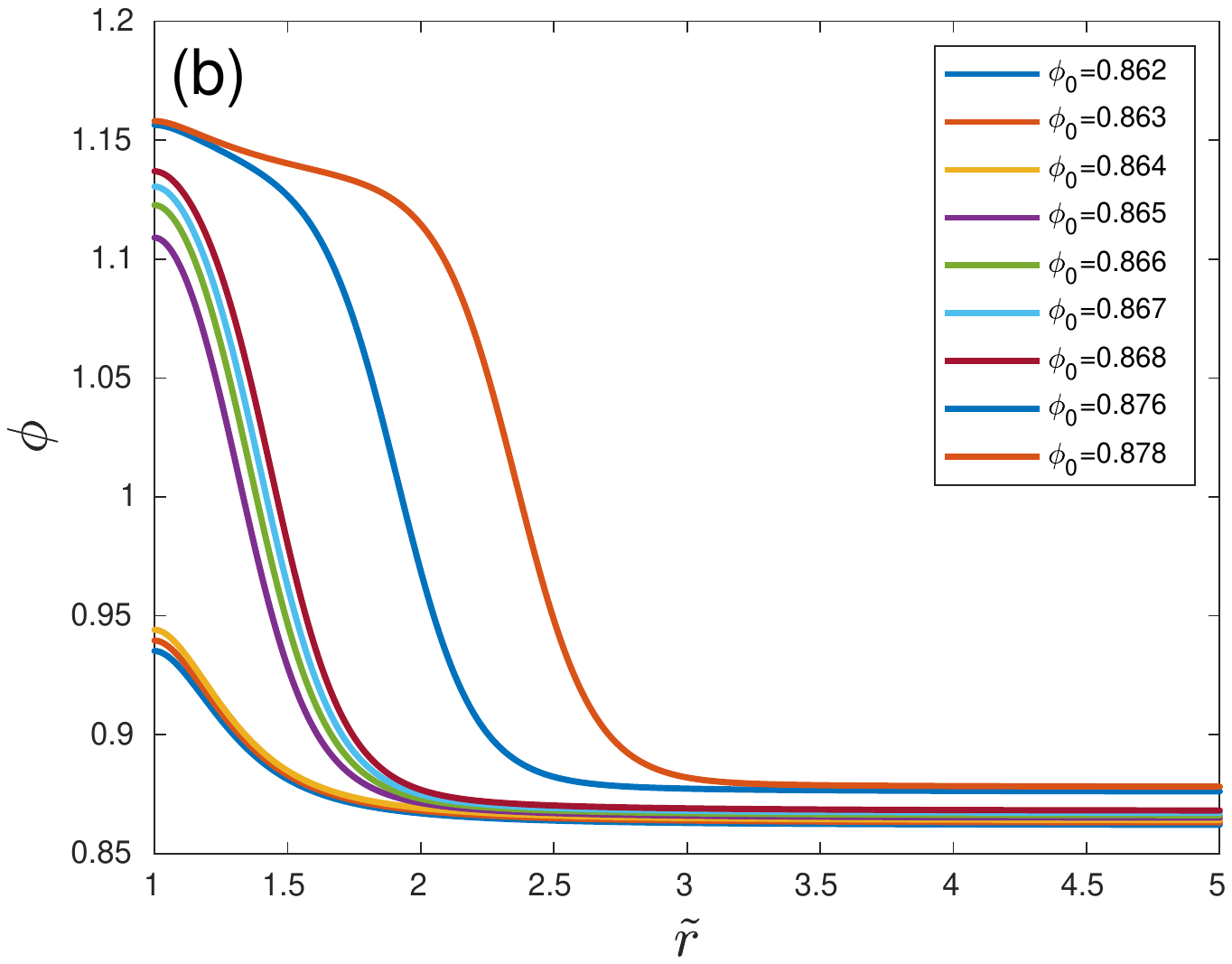}
\caption{Density profiles for a diffuse interface when $T/T_c=0.996$, $M=0.1$, 
and $\tilde{c}^2=0.001$. The initial guess for the iterations is taken as the 
solution of the previous iteration where $\phi_0$ is increasing 
(a) or decreasing (b).}
\label{comsol_profile}
\end{figure}

The density profiles for the case where $\tilde{c}^2=10^{-3}$ are presented in 
Figs. \ref{comsol_profile}(a) and \ref{comsol_profile}(b). In contrast to the 
CNT model, here, the density of both liquid and vapor phases is spatially 
varying, and the interface between them is smooth. We computed profiles for 
different values of bulk composition $\phi_0$ in iterations. The initial guess 
for the profile of each value of $\phi_0$ is taken as the solution of the 
previous iteration. Figures \ref{comsol_profile}(a) and \ref{comsol_profile}(b) 
show the same values of $M$, $T/T_c$, and 
$\phi_0$, but in Fig. \ref{comsol_profile}(a), $\phi_0$ was increased in the 
iterations from  unsaturated conditions to supersaturation, while in Fig. 
\ref{comsol_profile}(b), $\phi_0$ was reduced. This procedure leads to 
differences in the profiles in the range $\phi_0=0.865-0.867$. In this range, 
the system is in a metastable state in Fig. \ref{comsol_profile}(a). The 
energies of the profiles in Fig. \ref{comsol_profile}(b) are lower, indicating that
these are the equilibrium solutions for these conditions.

\begin{figure}[h!]
\includegraphics[width=0.48\textwidth,bb=95 235 495 545,clip]{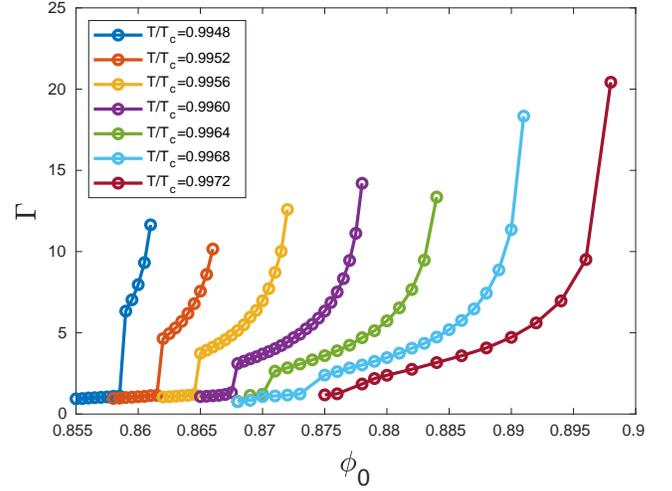}
\caption{Adsorption $\Gamma$ from Eq. (\ref{adsorption_eq}) vs bulk 
composition at different temperatures (see the legend). For a particular 
temperature, $\Gamma$ increases with $\phi_0$ for sufficiently small values of 
$\phi_0$. At the first-order transition line, it ``jumps'' discontinuously to 
higher values. Above this threshold composition, $\Gamma$ continuous to 
increase with $\phi_0$. It diverges at the binodal composition. For 
sufficiently high temperatures, given by Eq. (\ref{kink_point}), the 
first-order line becomes second-order, and the transition becomes smooth 
($T/T_c=0.9972$). We used $M=0.1$ and $\tilde{c}^2=0.001$.
}
\label{adsorption}
\end{figure}

Once profiles are found, one can calculate the adsorption $\Gamma$, given by 
the following expression:
\begin{eqnarray}\label{adsorption_eq}
\Gamma=4\pi\int_1^\infty 
\left(\phi(\tilde{r})-\phi_0\right)\tilde{r}^2d\tilde{r}.
\end{eqnarray}
The adsorption is shown in Fig. \ref{adsorption} as a function of bulk 
composition $\phi_0$ and for several temperatures (see the legend 
of Fig. \ref{adsorption}). At temperatures 
sufficiently lower than $T_c$, $\Gamma(\phi_0)$ increases slowly with $\phi_0$ 
until a certain critical composition is reached. At this composition, $\Gamma$ 
rises up sharply, which is the signature of the first-order phase transition 
(due to the deficiency of numerical schemes, the discontinuous jump has a finite 
width). Above the transition, $\Gamma$ continues to increase until, when 
$\phi_0$ equals the binodal composition at that particular temperature 
(the right-most point in every curve), it diverges. This behavior is reminiscent of 
the classical prewetting transition near a surface with short-range 
interactions. \cite{cahn_jcp_1977,pgg_rmp_1985,bonn_ross_rpp_2001} At 
temperatures sufficiently close to $T_c$, however, the transition is a 
continuous second-order transition, as should be expected in the presence of 
long-range forces.

\subsection {Phase diagram}\label{sec_Phase_Diagram_c_0}

In the absence of a charged particle, the phase equilibrium is given by the 
following classical common-tangent construction:
\begin{eqnarray}
\frac{\partial f_{\rm vdw}(\phi_1)}{\partial \phi}-\mu=0,\nn\\
\frac{\partial f_{\rm vdw}(\phi_2)}{\partial \phi}-\mu=0,\nn\\
\frac{f_{\rm vdw}(\phi_2)-f_{\rm vdw}(\phi_1)}{\phi_2-\phi_1}-\mu=0.
\end{eqnarray}
Here, $\phi_1$ and $\phi_2$ are the two coexisting binodal densities for the given 
temperature. The spinodal curve, defined by $\partial^2 f_{\rm vdw}/\partial 
\phi^2=0$, is below the binodal. While under the spinodal, phase separation is 
spontaneous, the area between the spinodal and binodal curves is metastable and 
liquid appears by nucleation. 

\begin{figure}[h!]
\includegraphics[width=0.48\textwidth,scale=0.65,bb=80 230 500 
540,clip]{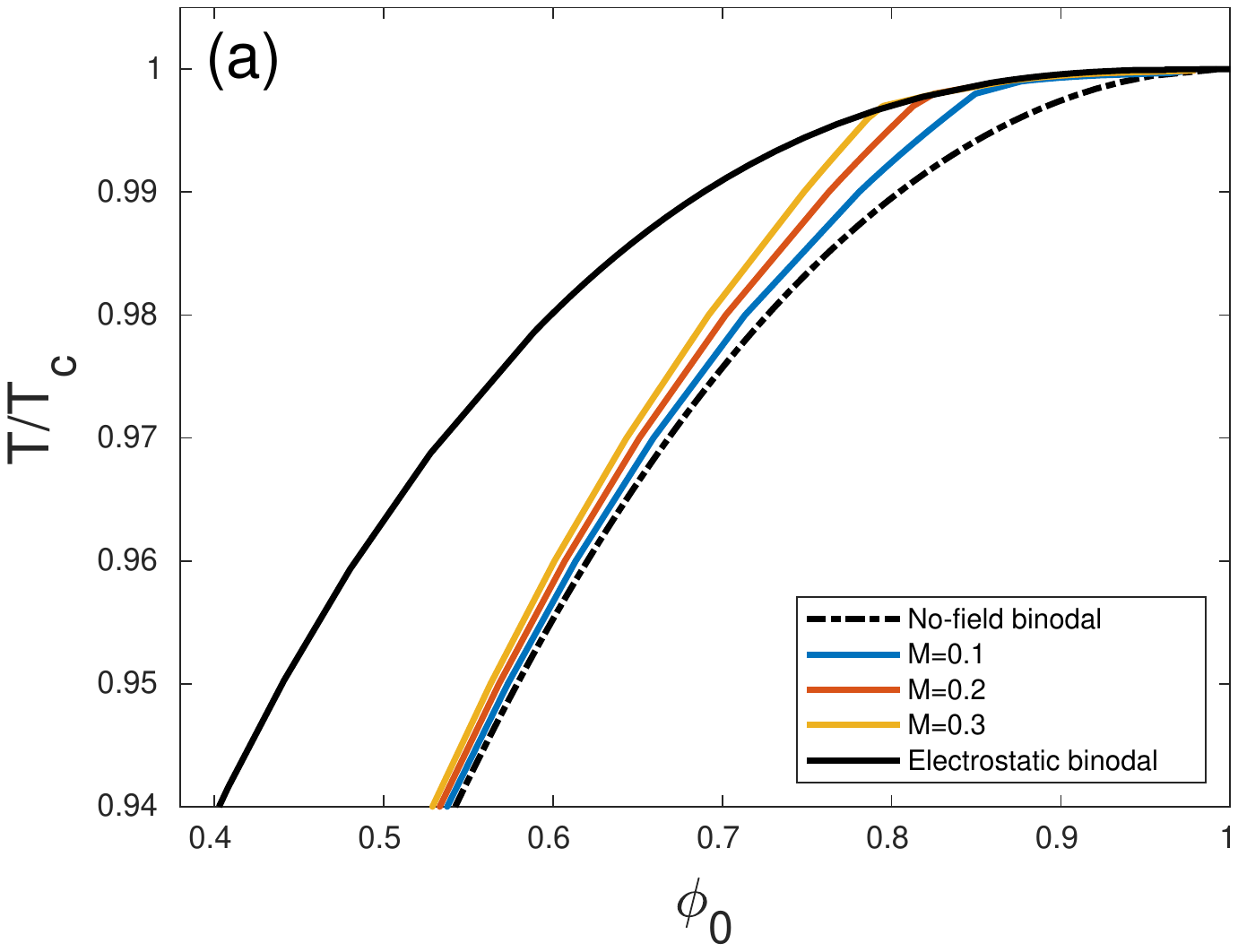}
\includegraphics[width=0.48\textwidth,scale=0.65,bb=80 230 500 
540,clip]{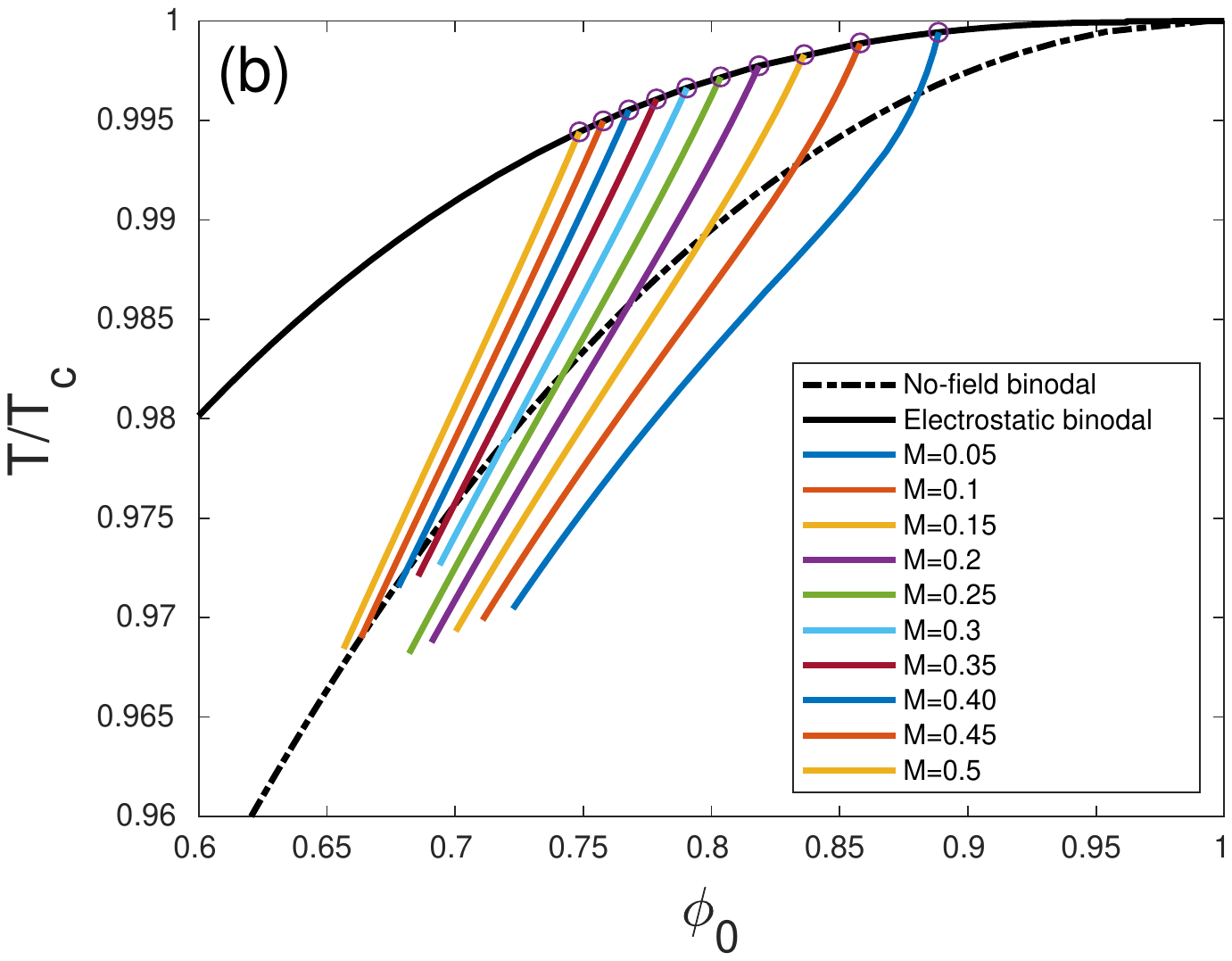}
\caption{(a) Stability diagram in the sharp interface limit. The black dashed-dotted 
line represents the classical no-field binodal curve. 
The black solid line is the electrostatic binodal. 
The colored lines represent the stability curves for different values of $M$. Under 
the stability curves, liquid--vapor coexistence is thermodynamically preferred. 
(b) Electrostatic spinodals in the sharp interface limit for different 
values of $M$. Under the 
electrostatic spinodal, the system spontaneously separates to two phases, while 
above it, it is in a metastable state.}
\label{Stability_diagram}
\end{figure}

The presence of a charged particle induces phase separation, and this modifies 
the phase diagram. Figure \ref{Stability_diagram}(a) shows the new phase diagram in 
the sharp interface limit. In the presence of a charged particle, we call the 
coexistence curve the ``stability curve.'' The stability curves for different 
values of $M$ appear as colored lines. Point $(\phi_0,T)$ above the stability 
curve is stable (the particle is surrounded by a vapor phase), while below this 
curve, the point is unstable and nucleation occurs (the particle is wetted by a 
dense phase). An increase in $M$ enlarges the range of conditions where 
nucleation occurs. While under the binodal, the radius of nucleation is 
infinite, in the area between the binodal and the stability curve, droplets with 
finite radius nucleate. The black solid line in Fig. \ref{Stability_diagram}
represents the ``electrostatic 
binodal.'' This curve represents the stability line for the limit 
$M\rightarrow\infty$. This means that for a point $(\phi_0,T)$ above the 
electrostatic binodal, there is no value of $M$ that can lead to nucleation. 
\cite{Galanis_2013} 

The critical point is modified in the presence of charge. 
The ``kink point'' is the point where the first-order transition line becomes a 
second-order transition. In the sharp interface limit, this point is defined by 
the pair of values $(\phi_0,T/T_c)$ that sets to zero the value of the second 
and third derivatives of $\tilde{f}$ at $\tilde{r}=1$,
\begin{eqnarray}
\frac{\partial \tilde{f}_{\rm vdw}(\phi_s)}{\partial 
\phi}-\frac{M\Delta\veps}{3\veps(\phi_s)^2}-\frac{\partial \tilde{f}_{\rm 
vdw}(\phi_0)}{\partial \phi}&=&0,\nn\\
\frac{\partial^2 \tilde{f}_{\rm vdw}(\phi_s)}{\partial 
\phi^2}+\frac{2}{9}
\frac{M\Delta\veps^2}{\veps(\phi_s)^3}&=&0,\nn\\
\frac{\partial^3 \tilde{f}_{\rm vdw}(\phi_s)}{\partial 
\phi^3}-\frac{2}{9}\frac{M\Delta\veps^3
}{\veps(\phi_s)^4}&=&0,
\label{kink_point}
\end{eqnarray}
where $\phi_{s}=\phi(\tilde{r}=1)$ is the surface composition.

In the reduced quantities used by us, the value of the bulk critical point 
without a charge is $(\phi_c,T/T_c)=(1,1)$. When $M$ is sufficiently small, the 
value of 
the kink point on the colloid surface can be found analytically 
by substituting $T/T_c=1+\Delta t$ and $\phi=1+\Delta\phi$ in Eq. 
(\ref{kink_point}) to get
\begin{eqnarray}
\Delta t\approx-\frac{M\Delta\veps^2}{27\veps_c^3},\nn\\
\Delta\phi\approx\frac{2M\Delta\veps^3}{81\veps_c^4},
\end{eqnarray}
where $\veps_c=\veps(\phi=1)=\veps_1+\Delta\veps/3$. 
The kink points for different values of $M$ are shown as circles in Fig. 
\ref{Stability_diagram}(b). 

The kink point is the end of another curve, the ``electrostatic spinodal.'' The 
electrostatic spinodal, in a manner similar to the regular non-field case, 
divides the two-phase region to the area where the spontaneous and non-spontaneous 
nucleation occurs, as shown in Sec. \ref{Diffuse_interface}. When 
$\tilde{c}=0$ and for a fixed temperature, it can be calculated analytically 
looking for the value of $\phi_0$ that leads to $\partial^2 
f(\tilde{r}=0)/\partial\phi^2=0$ at the surface of the colloid ($\tilde{r}=1$). 
For non-zero $\tilde{c}$, we find it by looking at the parameters that cause 
phase separation when the initial guess is the uniform vapor density $\phi_0$.  
 
\begin{figure}[h!]
\includegraphics[width=0.48\textwidth,scale=0.65,bb=55 205 515 
565,clip]{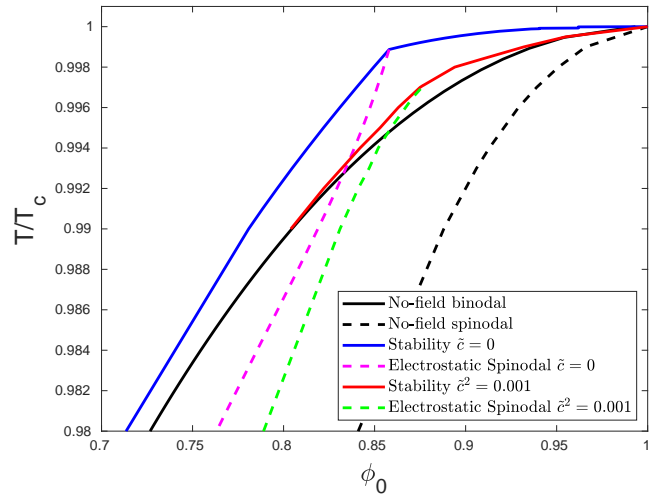}
\caption{Comparison between the stability curves for the 
sharp interface limit (blue) and the diffuse interface 
(red) with $M=0.1$. The corresponding electrostatic spinodals are in pink 
(sharp interface limit) and green (diffuse interface).}
\label{Zoomed_phase_diagram_c}
\end{figure}

Figure \ref{Zoomed_phase_diagram_c} shows the 
phase diagrams in both the sharp interface and diffuse interface limits.
While the electric charge enlarges the area where phase separation is 
favorable, the existence of the $(\nabla\phi)^2$ term leads to an effective
surface tension and to reduction in this area. 
Note that the change to the two-phase boundaries due to electric charge is 
appreciable relatively close to $T_c$, while the change in the electrostatic 
spinodal is significant at all temperatures. 
This can have an impact on supersaturated processes far from the critical 
temperature such as aerosol creation in clouds. \cite{Curtius2006}  

\section {CONCLUSIONS}

We study the vapor--liquid nucleation around a charged spherical particle using a 
simple mean-field approach. Electric field gradients lead to spatial 
nonuniformities in both the liquid and vapor phases, and this significantly 
alters the nucleation conditions. We find a first-order phase transition 
line outside of the binodal curve, similar to a prewetting line. Due to the 
long-range nature of electrostatic forces, this line becomes a second-order line 
at a special ``kink'' point, whose location is given approximately by Eq. 
(\ref{kink_point}).  The square-gradient approach yields diffuse-interface 
profiles that are more realistic than the CNT profiles and facilitates the 
calculation of the surface tension. For small particle and nucleus, the width of 
the interface can be appreciable and deviations from the CNT theory are 
significant. This is especially relevant close to $T_c$ where the wetting is 
continuous and energy barriers can be small. From the profiles, we construct a 
phase diagram, including the electrostatic binodal, spinodal, and kink point, 
indicating changes in the two-phase equilibrium region close to $T_c$. The 
existence of a stable liquid nucleus with a finite radius is predicted even 
outside of the binodal curve. 

We treated purely dielectric fluids. A question arises as to the effect of 
ionic screening. When the Debye lengths of the liquid and vapor phases are much 
smaller than the particle radius $R$, the field is localized at an exponentially 
thin layer at the surface of the colloid. In this limit, on large scales, the 
electrostatic energy should lead to an effective surface tension between the 
liquid and the surface. In the opposite limit, where
the Debye lengths are  much larger than $R$, one 
retrieves the case of dielectric liquids with the field decaying as $\sim 1/r^2$. 
The interesting case is the very large intermediate regime. 

Further investigation of the dynamics of the nucleation process is required. 
Such studies can increase the accuracy of predictions of the kinetics of
nuclei creation, and this may have implications in engineering applications and 
in atmospheric studies.

{\bf Acknowledgement} 

This work was supported by the Israel Science 
Foundation (ISF) grant No. 274/19.

{\bf Conflict of Interest}

We have no conflicts of interest to disclose.

{\bf Data Availability Statement}

The data that supports the findings of this study are available within the 
article [and its supplementary material].

\nocite{*}
\bibliography{roni_article2.bib}

\end{document}